\documentclass[12pt,a4paper]{article}

\usepackage{t1enc}
\usepackage[dvips,final]{graphicx} %<--ked chces vkladat obrazky
\usepackage{tabularx} %<--ked chces robit nieco s tabulkami
\usepackage{amsfonts}
\usepackage{amssymb,amsmath}

\def\du{\unskip\smash{\lower 1.4ex \hbox{\char34}}\kern-.2ex}
\def\hu{\kern-.2ex\hbox{\char92}}

\def\XXint#1#2#3{{\setbox0=\hbox{$#1{#2#3}{\int}$}
     \vcenter{\hbox{$#2#3$}}\kern-.5\wd0}}

\newcommand{\bdis}{\begin{displaymath}}
\newcommand{\edis}{\end{displaymath}}
\newcommand{\be}{\begin{equation}}
\newcommand{\ee}{\end{equation}}
\newcommand{\mbb}{\mathbb}

\newcommand{\pd}{\partial}
\newcommand{\noi}{\noindent}

\newtheorem{pr*}[thm]{*}

\begin{document}
\baselineskip=6mm
%\newpage

\title{A Toy Model for Black Hole in Noncommutative Spaces}
\author{Michal Demetrian\footnote{e-mail: demetrian@fmph.uniba.sk}, Peter Pre\v snajder\footnote{e-mail: presnajder@fmph.uniba.sk} \\
Comenius University, Mlynska Dolina 842 48 \\ Bratislava IV, Slovakia}

\maketitle

\abstract{ We present a new point of view on the problem of the Schwarzschild black hole in the
noncommutative spaces, proposed recently by F. Nasseri. We apply our treatment also to the case of
the 2+1 dimensional Ba\~ nados-Teitelboim-Zanelli black hole.}  \\
\noi

\section{Introduction}

Noncommutative geometry, \cite{connes}, and its physical applications have been studied extensively from various
points of view during last couple of years. The idea to study
the physical applications of the noncommutative geometry has
very good motivation in the following:
the dynamical variable in the general theory of relativity is the spacetime
itself, and we know that the procedure of quatization transfers the classical commutative dynamical variables into
the noncommuting operators in the quantum theory. Therefore, on can expect that the physics on the Planck scale is to be
described within the noncommutative spacetime. Moreover, this geometry arises naturally in the string theory, where the
noncommutativity is related to a kind of background field (magnetic-like field), \cite{sw},\cite{cds}. \\
It is very natural to suppose the black holes are the objects where the strong gravitational field meets with
effects of quantum mechanics and therefore the idea of a black hole in a noncommutative space is surely interesting. \\
In the section 2 of the paper we present a toy model of the Schwarzschild black hole in the noncommutative space and
the section 3 discuss a kind of 2+1 dimensional black hole.

\section{Incorporation of a space-space noncommutativity into the Schwarzschild metric}

The Schwarzschild metric is given by the line element
\be
\label{schwm} {\rm d}s^2=\left( 1-\frac{2GM}{r}\right){\rm
d}t^2-\left( 1-\frac{2GM}{r}\right)^{-1}{\rm d}r^2-r^2{\rm d}\Omega_2^2 .
\ee
In this metric, the sphere with the radius $r=2GM$ corresponds to the event horizon and
is given by the mass $M$ of the black hole only. The idea of the
paper \cite{nasseri} and some other recent papers like
\cite{nasseri1}, is to change the used coordinates into the
noncommuting ones in a manner many times used in the formulation
of quantum mechanics on noncommutative spaces
\cite{sheikh},\cite{belluci},\cite{berto},\cite{yin} and
\cite{jadenis}. However, in our problem this treatment can hardly
be used: how do we understand the objects like ${\rm d}\hat{r}$
with $\hat{r}$ being an operator? And how to understand the quantities like angular momentum and linear momentum
of undefined object(s) (bodies)
entering the results of the paper \cite{nasseri}? \\

We propose the following
modification of the treatment proposed by F. Nasseri. We
express the formula for the gravitational radius of the black hole
as the constraint:
\be \label{constr} r^2=4G^2M^2 \equiv r_g^2.
\ee
This constraint can be "quantized" introducing the
noncommuting variables $\hat{x}_i$, $i=1,2$ (in fact, we are going
to be interested in a planar section of the space only, nothing
more can be done within this concept of noncommutativity, more appropriate example will be discussed in the next section):
\be \label{cc1}
[\hat{x}_1,\hat{x}_2]=\theta , \quad \mbox{or} \quad
[\hat{x}_i,\hat{x}_j]=\theta \epsilon_{ij} ,
\ee
where the
parameter $\theta$ of dimension $length^2$ describes the
space-space noncommutativity and the square root of $|\theta|$ plays the role of a fundamental length $l_P$.
%We also add the commutation relations
%\bdis
%[\hat{x}_i,\hat{p}_j]=\delta_{ij},\quad [\hat{p}_i,\hat{p}_j]=0 ,
%\edis
%where the $\hat{p}$-s are the standard linear momentum
%operators.
Making the change
\be \label{change} r^2\mapsto
\hat{r}^2\equiv \hat{x}_1\hat{x}_1+\hat{x}_2\hat{x}_2
\ee
we obtain the operator version of the eq. (\ref{constr})
\be \label{constr1}
\hat{x}_1\hat{x}_1+\hat{x}_2\hat{x}_2= r_g^2 .
\ee
The commutation relations (\ref{cc1}) can be realized in terms of differential ope\-rators as follows:
\be \label{cc2}
\hat{x}_i=x_i-\frac{i}{2}\theta\epsilon_{ij}\partial_j , \quad i,j=1,2 \ ,
\ee
acting on auxiliary function $\psi(x)$. The constraint equations (\ref{constr1}) can be then represented as
the following eigenvalue problem (supposing $\psi$ belongs to $L^2(\mbb{R}^2)$):
\be \label{evp}
\left( x_i-\frac{1}{2\hbar}\theta\epsilon_{ij}p_j\right)
\left( x_i-\frac{1}{2\hbar}\theta\epsilon_{ik}p_k\right)\psi=r_g^2\psi .
\ee
Performing some algebra we derive the partial differential equation for $\psi$
\begin{eqnarray} \label{lleqnc}
& &
\left\{-\frac{\hbar^2}{2m}(\pd^2_1+\pd^2_2)+\frac{\hbar^2}{2m}\frac{4}{\theta^2}(x_1^2+x_2^2)+\frac{\hbar^2}{2m}
\frac{4i}{\theta}(x_1\pd_2-x_2\pd_1)\right\}\psi \nonumber \\
& &
=\frac{1}{2m}\frac{4\hbar^2 r_g^2}{\theta^2}\psi ,
\end{eqnarray}
where we introduced a formal, say positive, parameter $m$ to be able to compare easily
our equation with the Schr\" odinger
equation for the motion of a particle with mass $m$ and electric charge $e$ in the homogeneous magnetic field
$\vec{B}=(0,0,B)$ that reads
\begin{eqnarray} \label{lleq}
& &
\left\{ -\frac{\hbar^2}{2m}(\pd_1^2+\pd_2^2)+\frac{1}{2m}\frac{e^2B^2}{4}(x_1^2+x_2^2)+\frac{ieB\hbar}{2m}
(x_1\pd_2-x_2\pd_1)\right\}\psi  \nonumber \\
& &
=E\psi .
\end{eqnarray}
Comparing relevant terms of the eqs. (\ref{lleqnc}) and (\ref{lleq}) we identify
\bdis
eB=\frac{4\hbar}{\theta} ,
\edis
and having in mind the energies of the Landau levels: $E_n=\frac{\hbar eB}{m}(n+1/2)$,
we finally find the possible values of the black hole radius
$r_g$:
\be \label{qrrg}
r_g=\sqrt{2\theta\left( n+\frac{1}{2}\right)} ,
\ee
corresponding to the eigenfunction $\Phi_n$.
This is the quantization rule for the gravitational radius in our simple model of the Schwarzschild black hole in the
noncommutative space. \\

In this particular situation, we have obtained
the main result expected from the noncommutative geometry in physics - the quantization of a classical space-time
characteristic. Moreover, we see that our result is compatible with the well-settled suggestion that the
area of the event horizon of the Schwarzschild black hole $A$ is to be quantized as follows, \cite{qa}
\be \label{qaf}
A\equiv A_n=4\pi r_g^2=8\pi\theta\left( n+\frac{1}{2}\right)={\mathcal A}(n+\mathcal{O}(1))l_P^2 ,
\ee
where ${\mathcal A}$ is a dimensionless constant,
$l_P$ represents the Planck length and $n$ runs over all the natural numbers.
Obviously, our treatment the problem is not unique, and one can imagine other ways how to
consider the Schwarzschild black hole in a noncommutative space. For example, in the paper \cite{dolan}
the same problem is attacked by considering the event horizon of the black hole by the
well known fuzzy sphere, \cite{madore}, leading to the same kind of result for the quantization of the
black hole radius/mass is obtained.
We are not attempting to propose the presented approach as some roots for a definitive theory
but we would like to point out that the procedures presented in, say, \cite{nasseri} and
\cite{nasseri1}, may be improved. One of the possible improvements is our toy model,
an alternative is presented in \cite{dolan}.

\section{The case of Ba\~ nados-Teitelboim-Zanelli black hole}

In \cite{banados} Ba\~ nados,Teitelboim and Zanelli (BTZ)
have found the black-hole solution in 2+1 dimensional space-time with
negative cosmological constant. The line element reads
\begin{eqnarray} \label{bm}
& &
{\rm d}s^2=\left(-M+\frac{r^2}{l^2}+\frac{J^2}{4r^2}\right){\rm d}t^2-
\left(-M+\frac{r^2}{l^2}+\frac{J^2}{4r^2}\right)^{-1}{\rm d}r^2- \nonumber \\
& &
r^2\left({\rm d}\theta-\frac{J^2}{2r^2}{\rm d}t\right)^2,
\end{eqnarray}
where $M$ is the mass of the black hole, $\Lambda=-l^{-2}$ is the negative cosmological constant and $J$ is the
angular momentum of the black hole. The constraint
\bdis
-M+\frac{r^2}{l^2}+\frac{J^2}{4r^2}=0, \quad r^2=x_1^2+x_2^2
\edis
determines the event horizon of the black hole. \\

We can introduce noncommutativity into the $(x_1x_2)$ plane in the same manner as in the previous section.
In this way the condition for the horizon transforms into the form
\be \label{aohnc}
\left(M\hat{r}^2-\frac{1}{l^2}\hat{r}^4\right)\psi=\frac{J^2}{4}\psi  .
\ee
Using the results of previous section we have:
\be \label{qf2}
r^2=2\theta\left( n+\frac{1}{2}\right) , \quad \mbox{and} \quad
2\theta\left( n+\frac{1}{2}\right)M-\frac{1}{l^2}4\theta^2\left( n+\frac{1}{2}\right)^2=\frac{J^2}{4}
\ee
with the same eigenfunctions $\Phi_n$ as above.
Equation (\ref{qf2}) quantizes the values of the angular momentum for given $M$ and $l$ and tells
us that for given values of the parameters $M$ and $l$ of considered black hole, we
have only finite number of admissible values of the angular momentum $J$.
In the particular case $J=0$ the mass of the black hole is quantized according to the formula
\be \label{massq2}
M\equiv M_n=\frac{2\theta}{l^2}\left( n+\frac{1}{2}\right) .
\ee
We can compute the maximal value of the angular momentum for given values of $M$ and $l$. Using simple algebra,
we rewrite eq. (\ref{qf2}) into the form
\bdis
\frac{J^2}{4}=-\frac{4\theta^2}{l^2}\left[\left( n+\frac{1}{2}-\frac{Ml^2}{4\theta}\right)^2-
\frac{M^2l^4}{16\theta^2}\right]
\edis
from which we obtain the upper bound on black hole angular momentum:
\be \label{jb}
|J|\leq J_{max}= Ml .
\ee
The bound (\ref{jb}) is identical to the classical one when the two horizons of classical BTZ black hole merge.

\section{Discussion}

We have proposed a simple toy model of a black hole in a noncommutative space instead of the model proposed by
Nasseri in \cite{nasseri} and \cite{nasseri1} that mixes the field-theory and particle mechanics. Within our model
we were able to obtain quantization formula of the radius (mass) of the Schwarzschild black hole (\ref{qrrg}) and
the so-called Ba\~ nados-Teitelboim-Zanelli black hole (\ref{massq2}). We have shown that the upper
limit for the value of the modulus of the angular momentum of the BTZ black hole within our model
is the same as for the classical BTZ black hole, \cite{banados}.
Our result (\ref{qrrg}) coincides with
the result of the paper \cite{dolan} where the author modeled the quantum black hole by modeling its event
horizon as the noncommutative two dimensional sphere.

\subsubsection{Acknowledgement}
This work was supported by the project no. VEGA 1/3042/06.

\end{document}